\documentclass[12pt]{article}

\catcode`\@=11
\@addtoreset{equation}{section}

\global\arraycolsep=2pt
\usepackage{mathtools,mathrsfs, amsbsy, amssymb, latexsym, amsfonts, amscd, amsmath}
\usepackage[nosort]{cite}
\usepackage{graphicx}
\usepackage[usenames, dvipsnames]{color}

\usepackage[normalem]{ulem}
\usepackage{pdfpages}

\allowdisplaybreaks

\newcommand{\mL}{\mathcal L}

\definecolor{myGray}{rgb}{.7,.7,.7}
\newcommand{\gray}[0]{\color{myGray}}

\newcommand{\w}{\widetilde}
\newcommand{\txt}{\textrm}
\newcommand{\al}{\alpha}
\newcommand{\ga}{\gamma}
\newcommand{\e}{\textrm{e}}
\newcommand{\J}{\mathfrak{J}}
\newcommand{\Q}{\mathfrak{Q}}
\newcommand{\1}{\underline{1}}
\newcommand{\2}{\underline{2}}
\newcommand{\3}{\underline{12}}
\def\pp{\mathsf{p}}
\def\qq{\mathsf{q}}

\newcommand{\g}{\mathfrak{g}}
\newcommand{\gh}{\hat{\mathfrak{g}}}

\def\beq{\begin{equation}}
\def\eeq{\end{equation}}
\def\beqz{\begin{equation*}}
\def\eeqz{\end{equation*}}
\def\bea{\begin{eqnarray}}
\def\eea{\end{eqnarray}}

\def\ha{\mbox{\small $\frac{1}{2}$}}

\begin{document}

\begin{center}
\vspace*{2em}
{\large\bf  Affine $q$-deformed symmetry \\ and the classical Yang-Baxter $\sigma$-model}\\
\vspace{1.5em}
{
F. Delduc$^*$\footnote{E-mail:~francois.delduc@ens-lyon.fr},
T. Kameyama$^*$\footnote{E-mail:~takashi.kameyama@ens-lyon.fr},
M. Magro$^*$\footnote{E-mail:~marc.magro@ens-lyon.fr},
B. Vicedo$^\dagger$\footnote{E-mail:~benoit.vicedo@gmail.com}}
\end{center}

\vspace*{0.25cm}

\begin{center}
{*\it Univ Lyon, Ens de Lyon, Univ Claude Bernard, CNRS, Laboratoire de Physique,\\
F-69342 Lyon, France}
\\
{$^{\dagger}$\it School of Physics, Astronomy and Mathematics, University of Hertfordshire,\\
College Lane, Hatfield AL10 9AB, United Kingdom}
\end{center}

\bigskip

\begin{abstract}
The Yang-Baxter $\sigma$-model is an integrable deformation of the principal chiral model 
on a Lie group $G$.  The deformation breaks the $G \times G$ symmetry  
 to $U(1)^{\txt{rank}(G)} \times G$.  It is known that 
 there exist non-local conserved charges which, 
together with the unbroken $U(1)^{\txt{rank}(G)}$ local charges, form a Poisson algebra 
  $\mathscr U_q(\g)$, which is the semiclassical limit of the quantum group $U_q(\g)$, 
  with $\g$ the Lie algebra of $G$. 
  For a general Lie group $G$ with rank$(G)>1$, we extend the previous  result by   constructing 
local and non-local conserved charges satisfying all the defining relations of the
infinite-dimensional Poisson algebra 
$\mathscr U_q(L\g)$, the classical analogue of the quantum loop algebra 
$U_q(L\g)$, where $L\g$ is the loop algebra of $\g$. Quite unexpectedly, these 
defining relations are proved without encountering 
any ambiguity related to the non-ultralocality of this integrable $\sigma$-model. 
\end{abstract}

\setcounter{footnote}{0}
\setcounter{page}{0}
\thispagestyle{empty}


\newpage

\section{Introduction}

The Yang-Baxter $\sigma$-model is an integrable one-parameter deformation of the principal chiral model on any real Lie group constructed by C. {Klim$\check{\text{c}}$\'{\i}k} nearly fifteen years ago \cite{Klimcik:2002zj,Klimcik:2008eq}.
More recently, this model was rederived within the hamiltonian framework by means of deforming the integrable structure of the principal chiral model \cite{Delduc:2013fga}. In particular, it was shown that the Yang-Baxter $\sigma$-model belongs to a broader class of new integrable $\sigma$-models which have now come to be known as $\eta$-deformations. Soon after, the list of known integrable $\sigma$-models grew further still with the addition of the class of so-called $\lambda$-deformations introduced in \cite{Sfetsos:2013wia}. New deformations of both types were successively defined in
\cite{Klimcik:2002zj,Klimcik:2008eq,Delduc:2013fga,Sfetsos:2013wia,Delduc:2013qra,
Delduc:2014kha,Kawaguchi:2014qwa,Delduc:2014uaa,
Hollowood:2014rla,Hollowood:2014qma,Hoare:2014oua,
Sfetsos:2015nya}.

\medskip

One of the hallmarks of an integrable field theory is having an infinite-dimensional algebra of hidden symmetries. In this article we focus on such symmetries at the classical level. In the case of the principal chiral model on a Lie group $G$, for instance, there is an obvious global $G \times G$ symmetry corresponding to left and right multiplication of its $G$-valued field. The conserved charges associated with the left and right $G$-symmetries each combine with a tower of non-local charges to form classical analogues of the Yangian $Y(\g)$ \cite{MacKay:1992he,Bernard:1992ya}. Here we denote by $\g$ the Lie algebra of the Lie group $G$.
 
\medskip
 
The purpose of the present article is to identify the classical symmetry algebra of the Yang-Baxter $\sigma$-model. More precisely, we consider the inhomogenous Yang-Baxter $\sigma$-model, meaning that the skew-symmetric $R$-matrix which appears in the 
action of this field theory is the standard solution of the modified classical Yang-Baxter equation (mCYBE). In 
this setting, the global $G \times G$ symmetry of the principal chiral model is broken to $U(1)^{\txt{rank}(G)} \times G$ once the deformation parameter $\eta$ is switched on.

As in the undeformed case, the local charges associated with the right $G$-symmetry combine 
with non-local charges to form a classical analogue of the Yangian $Y(\g)$. This was shown 
in \cite{Kawaguchi:2010jg} for the Yang-Baxter model on $SU(2)$. This result can also be deduced 
from the analogous statement for a certain two-parameter deformation of the principal chiral 
model established in \cite{Itsios:2014vfa}
(see also \cite{Orlando:2010yh,Kawaguchi:2011mz,Orlando:2012hu,Kawaguchi:2013gma} 
in the $SU(2)$ case). From now on we shall no longer be concerned with this part of the infinite-dimensional symmetry algebra.

It was shown in \cite{Delduc:2013fga} that the left $G$-symmetry of the Yang-Baxter 
$\sigma$-model is $q$-deformed. That is to say, there exist non-local conserved charges which, 
together with the unbroken $U(1)^{\txt{rank}(G)}$ local charges, form a Poisson algebra 
denoted $\mathscr U_q(\g)$, the semiclassical limit of the quantum group $U_q(\g)$ (see for instance \cite{Ballesteros_2009}), where $q$ is a function of the deformation parameter $\eta$. This property generalises
what was first shown to hold in the $SU(2)$ case \cite{Kawaguchi:2011pf}.  It is natural to 
expect that the classical analogue of the Yangian $Y(\g)$ associated with the left 
$G$-symmetry of the principal chiral model should itself be deformed when $\eta$ is switched on. Specifically, one expects that the Poisson algebra $\mathscr U_q(\g)$ is enlarged to an 
infinite-dimensional Poisson algebra $\mathscr U_q(L\g)$, the classical analogue of the quantum loop algebra $U_q(L\g)$, where $L\g$ is the loop algebra of $\g$. This was indeed
shown to be the case when $\g = \mathfrak{su}(2)$ in 
\cite{Kawaguchi:2012gp,Kawaguchi:2012ve} and also when a Wess-Zumino term is present in \cite{Kawaguchi:2013gma}. 

\medskip

In this article we consider a general Lie group $G$ with rank$(G)>1$ and construct
local and non-local conserved charges satisfying all the defining relations of the
infinite-dimensional Poisson algebra $\mathscr U_q(L\g)$ (see \emph{e.g.} \cite{tolstoy_1992} for the defining relations of the quantum affine algebra $U_q(\gh)$ of which $U_q(L\g)$ is a quotient). The fact that this is possible is somewhat surprising. Indeed, Poisson brackets of generic non-local conserved charges in a classical integrable field theory are known to be ill-defined due to the presence of non-ultralocal terms in the Poisson brackets of the Lax matrix with itself \cite{Maillet:1985ec,Maillet:1985ek}. In fact, all proofs of the defining relations of the classical analogue of Drinfeld's first realization of the Yangian are in some sense incomplete since they require dealing with such ambiguities \cite{MacKay:1992he,MacKay:2004tc,Itsios:2014vfa}. It is worth stressing that, in this context, ambiguities are in fact already encountered when computing Poisson brackets of the level $0$ charges!
Furthermore, it was observed in \cite{Kawaguchi:2012gp,Kawaguchi:2012ve} that ambiguities 
also appear in the $SU(2)$ Yang-Baxter $\sigma$-model when deriving the $q$-Poisson-Serre 
relations of $\mathscr U_q(L \, \mathfrak{su}(2))$ involving the non-local charge associated 
with the affine root.
 As we shall see, the $SU(2)$ case appears not to be representative of the general situation. 
 There is no need to use any regularisation prescription when considering the defining 
 relations for higher rank cases. Note that there is also no ambiguity 
 for anisotropic $SU(2)$ Landau-Lifshitz $\sigma$-models \cite{Kameyama:2014bua}.

\medskip

Let us summarise the method we shall use to establish the defining relations of $\mathscr U_q(L\g)$. Following the analysis of \cite{Delduc:2013fga}, conserved charges are extracted from the monodromy $T^g(\lambda)$ of the gauge transformation $\mathcal L^g(\lambda, x)$ of the Lax matrix by the $G$-valued field $g$ of the model. Here we denote by $\lambda$ the spectral parameter and $x$ is the spatial coordinate. A
special role is played by the two poles at $\pm i \eta$ of the twist function \cite{Delduc:2013fga}. For these values of the spectral parameter, $\mathcal{L}^g(\pm i \eta,x)$ belong to opposite Borel subalgebras of the complexification of $\g$. This enables, in particular, to define a set of conserved charges $Q^E_{\pm \alpha_i}$ and $Q^H_{\alpha_i}$ associated with every simple root $\alpha_i$, $i = 1, \ldots, \text{rank}(G)$ of $\g$. The charges $Q^E_{\pm \alpha_i}$ are non-local whereas the charges $Q^H_{\alpha_i}$ associated with the unbroken $U(1)^{\txt{rank}(G)}$ symmetry are local. It was shown in \cite{Delduc:2013fga} that these charges satisfy the defining relations of $\mathscr U_q(\g)$. Moreover, the Cartan-Weyl basis of $\mathscr U_q(\g)$ is obtained by taking $q$-Poisson brackets of the generating charges $Q^E_{\pm \alpha_i}$ and $Q^H_{\alpha_i}$. This complete basis of non-local charges encoded in the two monodromy matrices $T^g(\pm i \eta)$ are schematically represented by the two halves of the middle line in Figure \ref{fig-1}.
 
Constructing conserved charges which satisfy the defining relations of the Poisson 
algebra $\mathscr U_q(L\g)$ requires going to the next order in the expansion of the 
monodromy $T^g(\lambda)$ at the points $\pm i \eta$. More generally, 
the order in this expansion corresponds to the level of the charges. For the purpose of describing the defining relations we only need two extra conserved charges, which we will call $\w{Q}^E_{\mp\theta}$, associated with the affine simple root $\alpha_0 = \delta - \theta$ and $-\alpha_0$, where $\theta$ is the highest root of $\g$ and $\delta$ is the imaginary root of $L\g$. These will be constructed from the coefficient of the generators $E^{\mp \theta}$ in the linear term of the expansion of the monodromy
$T^g(\lambda)$ around the point $\pm i \eta$, respectively
(see equation  \eqref{v1}). These are again depicted schematically at levels $\pm 1$ in Figure \ref{fig-1}.
We proceed to show that together with the level $0$ charges $Q^E_{\pm \alpha_i}$ and $Q^H_{\alpha_i}$ they satisfy the following Poisson bracket relations
\begin{subequations} \label{def-uq-affine}
\begin{align}
i\{Q^H_{\alpha_i},\w{Q}^E_{\pm\theta} \}&=
\pm d_i^{-1}(\theta,\al_i)\,\w{Q}_{\pm\theta}^E,
\\
i\{\w{Q}^E_{\theta},\w{Q}^E_{-\theta}\}&=\frac
{q^{d_\theta Q^H_{\theta}}-q^{-d_\theta Q^H_{\theta}}}{q^{d_\theta}-q^{-d_\theta}}\,,\\
i\{{Q}^E_{\pm\alpha_i},\w{Q}^E_{\pm\theta} \}&= 0,
\end{align}
where $Q^H_{\theta}$ is a certain linear combination of the
$Q^H_{\alpha_i}$ and, with $(\cdot,\cdot)$ denoting the inner product on the set of
roots of $\g$, we define $d_i = \ha (\alpha_i,\alpha_i)$ and $d_\theta = \ha (\theta, \theta)$.
Furthermore, $q=\e^\gamma$ and $\ga = - \eta/(1+\eta^2)^2$.
 Finally, we also prove the $q$-Poisson-Serre relations
\begin{align}
\{
\underbrace{Q^E_{\al_i},\{Q^E_{\al_i}, \cdots , \{Q^E_{\al_i}}_{\qq+1\,\txt{times}}
,\w{Q}^E_{-\theta}\}_{q} \cdots \}_{q}\}_{q}&=
0, \label{qPS1}\\
\{ \{ Q^E_{\al_i} ,\w{Q}^E_{-\theta} \}_{q},\w{Q}^E_{-\theta}
\}_{q}
&=
0, \label{qPS2}
\end{align}
\end{subequations}
where $\qq$ is the smallest positive integer such that $-\theta+(\qq+1)\al_i$ is not a root. Here the $q$-Poisson bracket of any pair of charges $A_\alpha$ and $A_\beta$ associated with roots $\alpha$ and $\beta$ is defined as
\beq
\{ A_\al,A_\beta \}_{q} =  \{
A_\al,A_\beta \}+i\ga\,(\al,\beta)A_\al A_\beta. \label{defqbracket}
\eeq
The above relations \eqref{def-uq-affine} together with the ones already proved in \cite{Delduc:2013fga} form the defining relations of the Poisson algebra $\mathscr U_q(L\g)$.
\setlength{\unitlength}{5cm}
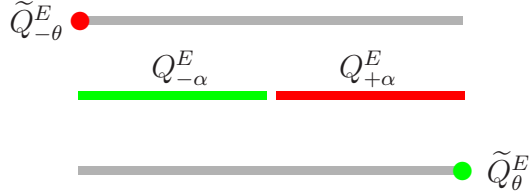
\begin{figure}[h]
\centering
\begin{picture}(1.6,.5)
\linethickness{1mm}
\put(.1,.37){$\w{Q}^E_{-\theta}$}
{\gray \put(.305,.4){\line(1,0){1}}}
{\color{red} \put(.26,.4){\circle*{.05}}}
\put(.45,.25){$Q^E_{-\alpha}$}\put(.95,.25){$Q^E_{+\alpha}$}
{\color{green} \put(.255,.2){\line(1,0){.5}}}
{\color{red} \put(.755,.2){\line(1,0){.5}}}
{\gray \put(.2,0){\line(1,0){1}}}
{\color{green} \put(1.195,0){\circle*{.05}}}\put(1.26,-.03){$\w{Q}^E_\theta$}
\end{picture}
\centering
\caption{The middle line depicts the level $0$ conserved charges, namely those of the finite-dimensional Poisson algebra $\mathscr U_q(\g)$, with the red and green portions corresponding to charges coming respectively
from $T^g( \pm i \eta)$. The dots on the ends of the upper and lower lines correspond
to two new level $\pm 1$ conserved charges of the infinite-dimensional Poisson algebra
$\mathscr U_q(L\g)$, coming respectively from the linear terms in the expansion of
$T^g(\lambda)$ around $ \pm i \eta$. } \label{fig-1}
\end{figure}

\section{The Yang-Baxter $\sigma$-model}

\paragraph{Action.}
The action of the Yang-Baxter $\sigma$-model is given by
\cite{Klimcik:2002zj,Klimcik:2008eq,Delduc:2013fga}
\begin{equation}
S=-\frac{1}{2}(1+\eta^2)^2\int \! dt \, dx
\kappa\Bigl(\partial_+g g^{-1}\,, \frac{1}{1-\eta R}\,\partial_-g g^{-1}\Bigr).
\label{YBsM}
\end{equation}
The field $g(t,x)$ takes values in a real semi-simple Lie group $G$ with Lie algebra
$\g$. We denote by $\g^{\mathbb{C}}$ the complexification of $\g$.
In this expression, $\kappa$ is the Killing form and
$\partial_\pm=\partial_t\pm\partial_x$. This integrable $\sigma$-model is characterised
by a skew-symmetric solution $R$ of the mCYBE. This means that the linear operator $R$ satisfies
\begin{subequations}
 \begin{align}
 \kappa(M,RN) &= - \kappa(RM,N),\\
  [RM,RN] &= R\bigl([RM,N]+[M,RN]\bigr) + [M,N]\label{mCYBE}
 \end{align}
\end{subequations}
for any $M,N\in \g$. Following \cite{Klimcik:2002zj,Klimcik:2008eq,Delduc:2013fga}, we
choose the standard  $R$-matrix of Drinfeld-Jimbo type
\cite{Drinfeld:1985rx,Jimbo:1985zk,Drinfeld:1986in}  (see equations \eqref{X} and \eqref{RX1} below).
The real
parameter $\eta$ plays the role of the
deformation parameter with $\eta=0$ corresponding to the principal chiral model. Finally, we
study the case when $x$ belongs to $\mathbb{R}$ and the field $g(t,x)$ tends
to constants when $x \to \pm \infty$.

\paragraph{Lax and monodromy matrices.}
The starting point is the Lax matrix $\mL^g(\lambda,x)$ defined in \cite{Delduc:2013fga}.
It takes values in
$\g^\mathbb{C}$ and has the following expression
\beq  \label{Lglambda}
\mathcal{L}^g(\lambda,x)= \frac{1}{1-\lambda^2}
\Bigl(
-(\lambda^2 + \eta^2)  \partial_xg(x) g^{-1}(x) + \frac{1}{1+\eta^2} (\lambda- \eta R) X(x)
\Bigr).
\eeq
The field $X(x)$ takes values in $\g$ and plays the role\footnote{More precisely,
$g(x)$ and $X(x)$ parameterise the cotangent bundle $T^\ast L G$ with $LG$ the loop group
associated with G.} of the field conjugate to $g(x)$ while $\lambda$ is the spectral parameter.
The central
object in our analysis of the symmetries of the Yang-Baxter $\sigma$-model is the monodromy
$T^g(\lambda)$, which is defined as the path ordered exponential of $\mathcal{L}^g(\lambda,x)$,
\beq
  T^g(\lambda) = P \overleftarrow{\exp}  \int_{-\infty}^\infty\!\!dx\,\mathcal{L}^g(\lambda,x).
\nonumber
\eeq
This monodromy is a conserved quantity, from which conserved charges will be extracted.

\paragraph{Fundamental Poisson brackets.}
The Poisson brackets of the fields $g(x)$ and $X(x)$ are
given by\cite{Delduc:2013fga}
\begin{subequations} \label{elementarypb}
\begin{eqnarray}
\{g_{\1}(x),g_{\2}(y)\}
&=&0\,,\label{gg}\\
\{X_{\1}(x),X_{\2}(y)\}
&=& [C_{\3},X_{\2}(x) ]\delta_{xy},\label{XX}\\
\{X_{\1}(x),g_{\2}(y) \}&=&C_{\3}\,g_{\2}(x)\delta_{xy},\label{Xg}
\end{eqnarray}
\end{subequations}
with $\delta_{xy}$ the Dirac distribution and
$C_{\3}$   the tensor Casimir. The definition of the latter
as well as notations used in \cite{Delduc:2013fga}
and in the present article are recalled in the next paragraph.

\paragraph{Notations.}

Given a choice of a Cartan subalgebra of the complexification $\g^\mathbb{C}$ we let $\Phi$ denote the associated root system. Let $\alpha_i$, $i=1,\cdots, n=\text{rk} \, \g^\mathbb{C}$ be a basis of simple roots,  and $E^{\pm \alpha}$ for $\alpha \in \Phi$ and $H^i = H^{\alpha_i}$ for $i = 1, \ldots, n$ be the corresponding Cartan-Weyl basis of $\g^\mathbb{C}$.
The matrix
\beq
B_{ij} =(\alpha_i,\alpha_j)= d_i A_{ij} \label{symCartanmatrix}
\eeq
is the symmetrised
Cartan matrix where $(\cdot,\cdot)$ is the inner product
on the set of roots and
\beq  \label{def di}
d_i = \ha (\alpha_i,\alpha_i).
\eeq
For any roots $\alpha,\beta \in \Phi$, we have
\beq \label{defofN}
[E^\alpha,E^\beta]= N_{\alpha,\beta} E^{\alpha +\beta}
\eeq
if $\alpha + \beta$ is a root, that is $\alpha + \beta \in \Phi$.  The Casimir tensor may then be written as
\beq
C_{\3}=\sum_{i,j=1}^n\,B_{ij}^{-1}\,H^i\otimes
H^j+\sum_{\al>0}(E^\al\otimes E^{-\al}+E^{-\al}\otimes E^{\al}).
\nonumber
\eeq
For later purposes, we express $X(x)$ and $\partial_x g(x) g^{-1}(x)$ as
\begin{subequations} \label{X and dgIg}
\begin{eqnarray}
X(x)&=&i\sum_{j=1}^n h_j(x)H^j +\frac{i}{2}\sum_{\alpha>0}\left(e_\alpha(x)
E^{\alpha} +e_{-\alpha}(x)E^{-\alpha}\right),
\label{X}\\
\partial_xg(x) g^{-1}(x)&=&i\sum_{j=1}^n J^H_{j}(x) H^j+\frac{i}{2}\sum_{\alpha>0}\left(
J_{\alpha}(x)E^{\alpha}+J_{-\alpha}(x)E^{-\alpha}
\right).\label{Jx}
\end{eqnarray}
\end{subequations}
Note that the $R$-operator acts on $X$ as
\beq
RX(x)=\ha\sum_{\alpha>0}\left(e_\alpha(x) E^{\alpha} -e_{-\alpha}(x)E^{-\alpha}\right). \label{RX1}
\eeq

\section{Expansion around the poles of the twist function} \label{sec: expansion}

In this section we expand the gauge-transformed monodromy matrix $T^g(\lambda)$ around the poles
$\pm i \eta$ of the twist function. We then recall how conserved
charges appearing in the defining relations of $\mathscr U_q(\g)$ are obtained at these poles.
Finally, we explain how the non-local charges needed to establish the defining relations
of $\mathscr U_q(L\g)$ are computed.

\subsection{Expansion of the Lax matrix}

We first expand the Lax matrix $\mathcal{L}^g(\lambda,x)$ defined
in \eqref{Lglambda} around the poles $\pm i \eta$
of the twist function. We obtain
\begin{eqnarray}
\mathcal{L}^g(\pm i \eta + \epsilon_\pm,x)
&=&-\frac{\eta}{(1+\eta^2)^2} \,(R\mp i)X(x)
\label{Lg}\\
&&+\frac{\epsilon_{\pm}}{1+\eta^2}\left[\frac{1-\eta^2\mp 2i\eta^2R}{(1+\eta^2)^2}X(x)
\mp 2i\eta\,\partial_x g(x) g^{-1}(x)
\right]+\mathcal{O}(\epsilon^2_\pm).\nonumber
\end{eqnarray}
By using the expressions \eqref{X and dgIg},
this can be rewritten as
\begin{eqnarray}
\mathcal{L}^g(\pm i \eta + \epsilon_\pm,x)&=&
\pm\ga\Bigl(\sum_{j=1}^n h_j(x)H^j+\sum_{\al>0}e_{\pm\al}(x)E^{\pm\al}\Bigr)\nonumber\\
&&+\frac{\epsilon_{\pm}}
{(1+\eta^2)^2}\Bigl(\frac{i}{2}\sum_{\alpha>0} \w{e}_{\mp\al}(x)E^{\mp\al}
+ \psi_\pm \Bigr)+\mathcal{O}(\epsilon^2_\pm),
\label{eLg}
\end{eqnarray}
where $\psi_\pm$ contain  terms proportional to generators $H^j$ and $E^{\pm \alpha}$ respectively.
Such terms will not play any role in establishing the defining relations of $\mathscr U_q(L\g)$.
We have also introduced the parameter
\beq
\ga = -\frac{\eta}{(1+\eta^2)^2},
\nonumber
\eeq
and defined
\begin{eqnarray}
\w{e}_{\pm\alpha}(x) =
e_{\pm\alpha}(x)\pm2i\eta(1+\eta^2)J_{\pm\alpha}(x).
\label{def.we}
\end{eqnarray}
Taking $\epsilon_\pm=0$ in \eqref{eLg}, we see that
$\mathcal{L}^g(\pm i \eta)$ belong to opposite Borel subalgebras of $\g^\mathbb{C}$.

\subsection{Expansion of the monodromy matrix}

To expand the monodromy matrix, we will apply the following identity
\begin{multline}
 P \overleftarrow{\exp}\biggl[\int_{-\infty}^{+\infty}\!\!dx\biggl(\sum_{i=1}^n
\partial_x\phi_i(x) H^i+\sum_{\al}L_{\al}(x) E^\al\biggr)\biggr]=\txt{exp}\biggl(
\sum_{i=1}^n\phi_i(+\infty)H^i\biggr) \times \\
 \times  P \overleftarrow{\exp}\biggl[\int_{-\infty}^{+\infty}\!\!dx
\sum_{\al}\e^{-\sum_{i=1}^n (\alpha, \alpha_i) \phi_i(x)}L_{\al}(x)E^{\al}\biggr]\txt{exp}\biggl(-
\sum_{i=1}^n\phi_i(-\infty)H^i\biggr) \label{Id}
\end{multline}
to specific functions $\phi_i(x)$ and $L_\alpha(x)$.
Firstly, to expand $T^g(\lambda)$ around $i \eta$ we start from \eqref{eLg} and choose,
for $\al>0$,
\beqz
\phi_i(x)=\int_{-\infty}^xdy\,\ga h_i(y), \qquad
L_{\alpha}(x) = \gamma e_{\alpha(x)},
\qquad
L_{-\alpha}(x) =\frac{\epsilon_+}{(1+\eta^2)^2}\,\frac{i}{2}\,\w{e}_{-\al}(x).
\eeqz
This leads to
\begin{subequations} \label{eTall}
\begin{multline}
T^g( i \eta + \epsilon_+) =
\e^{\gamma\sum_{i=1}^n\int_{-\infty}^\infty\!\!dx\, h_i(x)H^i
}
P \overleftarrow{\exp}\Biggl[\int_{-\infty}^\infty\!\!dx\, \biggl(\gamma\sum_{\alpha>0}
\J_\al^E(x)E^{\alpha} + \\
+\frac{\epsilon_{+}}{(1+\eta^2)^2}\biggl[
\frac{i}{2}\sum_{\al>0}\w{\J}^E_{-\al}(x)E^{-\al}
+ \tilde{\psi}_+
\biggr]+
\mathcal{O}(\epsilon^2_+)\biggr)
\Biggr].\label{eTp}
\end{multline}
where $\tilde{\psi}_+$ contains terms proportional to generators $H^j$ and $E^{\alpha}$ with $\al>0$.
The expressions of $ \J_\al^E(x)$ and $\w{\J}^E_{-\al}(x)$ will be given shortly.
Secondly, to expand $T^g(\lambda)$ around $- i \eta$ we start from \eqref{eLg} and choose, for $\al>0$,
\beqz
\phi_i(x)=\int^{\infty}_x dy\,\ga h_i(y) , \qquad
L_{\alpha}(x) = \frac{\epsilon_{-}}{(1+\eta^2)^2}\,\frac{i}{2}\,\w{e}_{\al}(x),
\qquad
L_{-\alpha}(x) =-\gamma e_{-\al}(x),
\eeqz
which leads to
\begin{multline}
T^g(-i \eta + \epsilon_-)= 
P \overleftarrow{\exp}
\Biggl[\int_{-\infty}^\infty\!\!dx\biggl(-\gamma\sum_{\alpha>0} \J^E_{-\alpha}(x)E^{-\alpha}+ \\
+
\frac{\epsilon_{-}}{(1+\eta^2)^2}\biggl[
\frac{i}{2}\sum_{\al>0}\w{\J}^E_{\al}(x)E^{\al}
+ \tilde{\psi}_- \biggr]
+ \mathcal{O}(\epsilon^2_-) \biggr)
\Biggr]
\e^{-\gamma\sum_{i=1}^n\int_{-\infty}^\infty\!\!dx\, h_i(x)H^i}. \label{eTm}
\end{multline}
\end{subequations}
In \eqref{eTall}, we have  introduced the following functions,
\beq
\mathfrak{J}_{\pm\alpha}^E(x) = \e^{\ga\chi_{\alpha}(\mp\infty)}\,
\e^{-\gamma\,\chi_{\alpha}(x)}\,e_{\pm\alpha}(x)  \qquad  \mbox{and} \qquad
\w{\mathfrak{J}}_{\pm\al}^E(x) = \e^{\ga\chi_{\al}(\mp\infty)}\,\e^{+\gamma\,\chi_{\al}(x)}\,
\w{e}_{\pm\al}(x),\label{defwJ}
\eeq
where $\chi_\alpha(x)$ is defined as
\beq
\chi_\alpha(x) = \ha \sum_{i=1}^n (\al,\al_i)  \int_{-\infty}^\infty\!\!dy \, \epsilon_{xy}
\,h_i(y).\nonumber
\eeq
Here the signature function is defined as $\epsilon_{xy}\equiv\theta_{xy}-\theta_{yx}$
and $\theta_{xy}=\theta(x-y)$ is the Heaviside step function.

\subsection{Defining relations of $\mathscr U_q(\g)$}

Let us recall the result of the analysis carried out in \cite{Delduc:2013fga}, the starting point of which is the limit $\epsilon_\pm=0$ of \eqref{eTall}. Making a choice of a normal ordering on the set of positive roots, \emph{i.e.} such that if $\alpha < \beta$ and
$\alpha + \beta$ is a root then  $\alpha < \alpha + \beta < \beta$, one can write $T^g(\pm i \eta)$ as
\begin{subequations} \label{Tallpoles}
\begin{eqnarray}
T^g(i \eta )&=&
\exp\Bigl(\ga\int_{-\infty}^\infty\!\!dx\sum_{i=1}^nh_i(x)H^i\Bigr)
\prod_{\al>0}^{<}\exp
\Bigl(\ga
\int_{-\infty}^\infty\!\!dx\,\Q_{\al}^E(x)
E^\al
\Bigr),\\
T^g( - i \eta)&=&
\prod_{\al>0}^{>}\exp
\Bigl(-\ga
\int_{-\infty}^\infty\!\!dx\,\Q_{-\al}^E(x)
E^{-\al}
\Bigr)\exp\Bigl(-\ga\int_{-\infty}^\infty\!\!dx\sum_{i=1}^n h_i(x)H^i\Bigr).
\end{eqnarray}
\end{subequations}
The superscripts $<$ and $>$ refer to the choice of normal ordering of positive and negative
roots respectively. It is easy to see that for simple roots we have $\Q^E_{\pm \al_i}(x) = \J^E_{\pm \al_i}(x)$, where the latter were defined in \eqref{defwJ}. The   conserved charges associated
with Cartan generators and simple roots $\alpha_i$
are then
\beq \label{QHQEalphai}
Q_{\alpha_i}^H = d_i^{-1} \int_{-\infty}^\infty\!\!dx \,\mathfrak{J}_{\alpha_i}^H(x) \qquad
\mbox{and} \qquad
Q_{\pm\alpha_i}^E = D_i\int_{-\infty}^\infty\!\!dx \,\mathfrak{J}_{\pm\alpha_i}^E(x).
\eeq
For the densities
associated with Cartan generators, we have
\beq \label{defJH2}
\mathfrak{J}_{\alpha_i}^H(x) = \sum_{j=1}^n B_{ij}\,h_j(x)
\eeq
with
\beq
D_i = \Bigl( \frac{\ga}{4\sinh(d_i\ga)}\Bigr)^\frac{1}{2}.\label{defD}
\eeq
The symmetrised Cartan matrix $B_{ij}$ and $d_i$ are
defined in \eqref{symCartanmatrix}   and \eqref{def di}, respectively.

The Poisson brackets of the charges \eqref{QHQEalphai} are then found to be
\begin{subequations} \label{PB not affine}
\begin{eqnarray}
i\{Q^H_{\alpha_i},Q^H_{\al_j}\}&=&0,
\\
i\{Q^H_{\alpha_i},Q^E_{\pm\al_j}\}&=&
\pm A_{ij}\,Q_{\pm \al_j}^E,
\\
i\{Q^E_{+\al_i},Q^E_{-\al_j}\}&=&\delta_{ij}\,\frac{q^{d_i Q^H_{\al_i}}-q^{-d_i Q^H_{\al_i}}}{q^{d_i}-q^{-d_i}},
\end{eqnarray}
\end{subequations}
where $q \in \mathbb{R}$ is related to the deformation parameter $\eta$ as $q= \e^\gamma$.
We refer to  \cite{Delduc:2013fga} for the statement and derivation of the $q$-Poisson-Serre relations.

\subsection{The $\al_i$-string through $-\theta$}

To prove the defining relations of $\mathscr U_q(L\g)$, we shall study
charges associated with the string of roots $-\theta + r \alpha_i$ with $r$ taking values
from 0 to the smallest strictly positive integer $\qq$ such that $-\theta+(\qq+1)\al_i$ is not a root. These roots are ordered as
\beqz
-\theta<-\theta+\al_i<-\theta+2\al_i<\ldots<-\theta+\qq\,\al_i.
\eeqz
Expanding the path-ordered exponential appearing in \eqref{eTp}, we may write it as
\begin{subequations} \label{expA}
\beq
T^g(i \eta + \epsilon_+) =
\e^{\ga\int_{-\infty}^\infty\!\!dx\sum_{i=1}^n h_i(x)H^i}
\Bigl(1+ \frac{i \epsilon_{+} }{2(1+\eta^2)^2}   v^+ + \mathcal{O}(\epsilon^2_+) \Bigr)
\prod_{\al>0}^{<}\e^{
\ga
\int_{-\infty}^\infty\!\!dx\,\Q_{\al}^E(x)
E^\al
 }.\label{v1}
\eeq
We are only interested in terms from $v^+$ which will contribute to the $\al_i$-string through $-\theta$ defined above. We write such terms as
\beq \label{314b}
\sum_{r=0}^{\mathbf{q}} \int_{-\infty}^\infty\!\!dx \,  \w{\Q}^E_{-\theta+ r \al_i}(x)E^{-\theta+ r \al_i}.
\eeq
\end{subequations}
This will be our definition of the charge densities $\w{\Q}^E_{-\theta+ r \al_i}(x)$.
 We proceed in the same way for $T^g(-i \eta + \epsilon_-)$.

\paragraph{Charge densities $\w{\Q}^E_{ \mp \theta}(x)$ and Charges $\w{Q}^E_{ \mp \theta}$.}

The simplest charge densities of interest can be obtained directly from \eqref{eTall}, namely we have
\beq
\w{\Q}^E_{ \mp \theta}(x)=\w{\J}_{ \mp \theta}^E(x),
\label{defwQ0}
\eeq
where $\w{\J}_{ \mp \theta}^E(x)$ were defined in \eqref{defwJ}.
 We also define the associated charges
\beq  \label{defChargeTheta}
\w{Q}^E_{ \mp \theta}=D_{\theta}\int_{-\infty}^\infty\!\!dx\,\w{\Q}^E_{\mp\theta}(x),
\eeq
where
\beq
d_\theta = \ha(\theta,\theta) \qquad \mbox{and} \qquad
D_\theta = \Bigl( \frac{\ga}{4\sinh(d_\theta\ga)}\Bigr)^\frac{1}{2}.
\nonumber
\eeq

\paragraph{Charge densities $\w{\Q}^E_{-\theta+r\al_i}(x)$ and Charges $\w{Q}^E_{-\theta+r\al_i}$.}
 
For completeness and to illustrate the mechanism behind the expansion \eqref{expA}, we also indicate briefly how the other charges would be computed.
We stress, however, that we do not strictly need this derivation to prove the defining relations.
Indeed, in the next section we shall obtain these charges recursively as
$q$-Poisson brackets of conserved charges. This ensures that they are
themselves conserved. In particular, we shall see later that the charge densities $\w{\Q}^E_{-\theta+r\al_i}(x)$ with $ 0<r\leq\qq$ are expressed recursively as
\beq 
\w{\Q}^E_{-\theta+r\al_i}(x)=\w{\J}^E_{-\theta+r\al_i}(x)-\ga
N_{-\theta+(r-1)\al_i,\al_i}\,\J_{\al_i}^E(x)\int_{-\infty}^x\!\!dy\,\w{\Q}^E_{-\theta+(r-1)\al_i}(y),
\label{defwQ1}
\eeq
with $N_{-\theta+(r-1)\al_i,\al_i}$ defined in \eqref{defofN}.
We define the associated charge\footnote{The normalisations in \eqref{defChargeTheta} and \eqref{defQstring} are
fixed for later convenience.}
\beq \label{defQstring}
\w{Q}^E_{-\theta+r\al_i}= D_i^r D_{\theta}\int_{-\infty}^\infty\!\!dx\,\w{\Q}^E_{-\theta+r\al_i}(x).
\eeq

Consider the case $r=1$. We see from the expression \eqref{v1}, taking into account
\eqref{314b}, that it contains terms
in $E^{-\theta+ \al_i}$ and $E^{-\theta} E^{\al_i}$ but no terms in $E^{\al_i} E^{-\theta}$.
On the other hand, by expanding \eqref{eTp} we would get
\begin{multline}
\int_{-\infty}^\infty\!\!dx\biggl[
\,\w{\J}_{-\theta+\al_i}^E(x)E^{-\theta+\al_i}
+\ga\,\w{\J}^E_{-\theta}(x)\int_{-\infty}^\infty\!\!dy\,
\J_{\al_i}^E(y)\,\theta_{xy}\,\,E^{-\theta}E^{\al_i} +\\
+\ga\,\J_{\al_i}^E(x)\int_{-\infty}^\infty\!\!dy\,
\w{\J}^E_{-\theta}(y)\,\theta_{xy}\,E^{\al_i}E^{-\theta}
\biggr].
\label{int}
\end{multline}
Yet using the relation $E^{\al_i}E^{-\theta}=-[E^{-\theta},E^{\al_i}]+E^{-\theta}E^{\al_i}$
where $[E^{-\theta},E^{\al_i}]=N_{-\theta,\al_i}E^{-\theta+\al_i}$\,, we may rewrite \eqref{int} as
\begin{multline}
\int_{-\infty}^\infty\!\!dx \Bigl(\w{\J}^E_{-\theta+\al_i}(x)-\ga
N_{-\theta,\al_i}\,\J_{\al_i}^E(x)\int_{-\infty}^x\!\!dy\,\w{\J}^E_{-\theta}(y)\Bigr)E^{-\theta+\al_i}
\nonumber\\
+\ga \int_{-\infty}^\infty\!\!dx \int_{-\infty}^\infty\!\!dy\Bigl(\,\J_{\al_i}^E(x)\w{\J}^E_{-\theta}(y)+
\J_{\al_i}^E(y)
\w{\J}^E_{-\theta}(x)
\Bigr)\theta_{xy}\,E^{-\theta}E^{\al_i} .
\end{multline}
The first line above allows us to identify $\w{\Q}^E_{-\theta+\al_i}(x)$ as in \eqref{defwQ1} while the term in the second line gives $\gamma Q^E_{\al_i} \w{Q}^E_{-\theta} E^{-\theta}E^{\al_i}$.

\section{Defining relations of $\mathscr U_q(L\g)$}
 
In this section we prove that the defining relations of $\mathscr U_q(L\g)$ are satisfied. 
The computations are straightforward but quite lengthy. For this reason, intermediate 
Poisson brackets have been collected in the Appendix. Let us stress that no 
ambiguity is encountered when proving these defining relations. We shall comment on 
this in the next section.
 
\subsection{First set of defining relations}

The first result concerns the Poisson brackets between the level $\pm 1$ charges $\w{Q}^E_{\pm\theta}$ introduced in \eqref{defChargeTheta} and the level $0$ charges \eqref{QHQEalphai}. Since the derivation closely follows that of \cite{Delduc:2013fga} we omit the details. Starting from the definitions 
\eqref{defwJ} and \eqref{defJH2}, we find, after some algebra
\begin{eqnarray*}
\{\mathfrak{J}^H_{\alpha_i}(x),\w{\mathfrak{J}}^E_{\pm\theta}(y)\}&=&\mp i\,(\theta,\al_i)\,\w{\mathfrak{J}}^E_{\pm\theta}(x)\delta_{xy},
\label{PJH-Jth}\\
\{\w{\mathfrak{J}}^E_{\theta}(x),\w{\mathfrak{J}}^E_{-\theta}(y)\}&=&
-4i \,\partial_x\chi_{\theta}(x)\e^{2\ga\chi_{\theta}(x)}\delta_{xy},
\label{PJth-Jth}\\
\{{\mathfrak{J}}^E_{\pm\alpha_i}(x),\w{\mathfrak{J}}^E_{\pm\theta}(y)\}&=&0.\label{PJal-Jth}
\end{eqnarray*}
This allows one to deduce the following Poisson brackets of conserved charges
\begin{subequations} \label{firstsetdr}
\begin{eqnarray}
i\{Q^H_{\alpha_i},\w{Q}^E_{\pm\theta}\}&=&
\pm d_i^{-1}(\theta,\al_i)\,\w{Q}_{\pm\theta}^E,
\\
i\{\w{Q}^E_{\theta},\w{Q}^E_{-\theta}\}&=&\frac{q^{d_\theta Q^H_{\theta}}-q^{-d_\theta Q^H_{\theta}}}{q^{d_\theta}-q^{-d_\theta}},
\label{PQt}\\
i\{{Q}^E_{\pm\alpha_i},\w{Q}^E_{\pm\theta}\}&=&0,
\end{eqnarray}
\end{subequations}
where the conserved charge $Q_\theta^H$ is defined as
\beqz
Q_\theta^H =  d_\theta^{-1}\int_{-\infty}^\infty\!\!dx\,\mathfrak{J}_\theta^H(x)
\qquad \mbox{with} \qquad
\mathfrak{J}_\theta^H(x) = \sum_{i=1}^n\,(\theta,\al_i)\,h_i(x),
\eeqz
and $h_i(x)$ are defined by \eqref{X}.
Note that the charge $Q_\theta^H$ is not independent of the Cartan charges
 $Q_{\al_i}^H$ since we have the linear relation
\begin{eqnarray}
d_\theta Q_\theta^H=\sum_{i,j=1}^nB_{ij}^{-1}\,(\theta,\al_i)\,d_jQ_{\al_j}^H.
\nonumber
\end{eqnarray}
The results \eqref{firstsetdr} are among the defining relations of $\mathscr U_q(L\g)$.

\subsection{$q$-Poisson-Serre relations}

We now turn to the proof of the $q$-Poisson-Serre relations
\begin{subequations}
\begin{align}
\{
\underbrace{Q^E_{\al_i},\{Q^E_{\al_i}, \cdots , \{Q^E_{\al_i}}_{\qq+1\,\txt{times}}
,\w{Q}^E_{-\theta}\}_{q} \cdots \}_{q}\}_{q}&=
0, \label{qPS1bis} \\
\{ \{ Q^E_{\al_i} ,\w{Q}^E_{-\theta} \}_{q},\w{Q}^E_{-\theta}
\}_{q}
&=
0. \label{qPS2bis}
\end{align}
\end{subequations}
Recall the definition of the $q$-Poisson bracket in \eqref{defqbracket} and of the relevant charge densities in \eqref{defwQ0} and \eqref{defwQ1}.
We shall need the following properties of $N_{\al,\beta}$ defined in \eqref{defofN}.
Consider the $\alpha$-string through $\beta$ whose roots are
$\beta+\pp\al,\ldots,\beta,\ldots,\beta+\qq\al$, with $\pp\leq0$ and $\qq\geq0$.
Then,
\begin{eqnarray}
N_{\al,\beta}^2=\qq(1-\pp)\frac{(\al,\al)}{2} \qquad \mbox{and} \qquad
\frac{2(\beta,\al)}{(\al,\al)}=-(\pp+\qq).
\label{rel.N}
\end{eqnarray}
For the $\al_i$-string through $-\theta$,
since $-\theta-\al_i$ is not a root,
we have
\begin{eqnarray}
\pp=0,\qquad
\qq=
\frac{2(\theta,\al_i)}{(\al_i,\al_i)},\qquad
N_{-\theta,\al_i}^2=N_{\al_i,-\theta}^2=(\theta,\al_i). \label{idsurN}
\end{eqnarray}
For any $r$ such that $0\leq r\leq\qq$, we then have the following identities
\begin{eqnarray}
N_{-\theta+r\al_i,\al_i}^2&=&N_{-\theta+(r-1)\al_i,\al_i}^2+(\theta-r\al_i,\al_i),\nonumber\\
N_{-\theta+r\al_i,\al_i}^2&=&\left((r+1)\theta-\frac{r(r+1)}{2}\,\al_i,\al_i\right).
\label{rel.N-theta}
\end{eqnarray}
We shall also need the identities for the step functions,
\begin{eqnarray}
\theta_{yy'}\theta_{xy'}&=&\theta_{yy'}\theta_{xy}+\theta_{xy'}\theta_{yx},\label{id.step}\\
\theta_{yy'}\theta_{y'x}&=&\theta_{yy'}\theta_{yx}-\theta_{xy'}\theta_{yx}\label{id.step1}.
\end{eqnarray}

\paragraph{$q$-Poisson bracket $\{Q^E_{\al_i},\w{Q}^E_{-\theta}\}_{q}$.}

Consider the Poisson bracket $\{\J_{\al_i}^E(x),\w{\J}^E_{-\theta}(y)\}$ given in \eqref{512a}. Rewriting
$\theta_{xy}=\frac{1}{2}(\epsilon_{xy}+1)$ and
using \eqref{idsurN}, we obtain the $q$-Poisson bracket
\beq
\{
\J^E_{\al_i}(x),\w{\J}^E_{-\theta}(y)\}_{q}=
2iN_{-\theta,\al_i}\left(\w{\J}^E_{-\theta+\al_i}(x)\,\delta_{xy}-\ga\,
N_{-\theta,\al_i}\J^E_{\al_i}(x)\w{\J}^E_{-\theta}(y)\,\theta_{xy}\right).\label{qP}
\eeq
Integrating (\ref{qP}) over $x$ and $y$ and using the definition
\eqref{defwQ1} for $r=1$, the $q$-Poisson bracket between $Q_{\al_i}^E$ and $\w{Q}^E_{-\theta}$ is found to be
\begin{eqnarray}
\{
Q^E_{\al_i},\w{Q}^E_{-\theta}\}_{q}
=2iN_{-\theta,\al_i}\w{Q}^E_{-\theta+\al_i}.\label{qPQ1}
\end{eqnarray}

\paragraph{$q$-Poisson-Serre relation \eqref{qPS1bis}.}

Next, we compute the $q$-Poisson bracket between $Q^E_{\al_i}$ and $\w{Q}^E_{-\theta+\al_i}$
in a similar way. Using the relations (\ref{PJJ}), the result for the Poisson bracket between
$\J^E_{\al_i}(x)$ and $\w{\mathfrak{Q}}^E_{-\theta+\al_i}(x)$ can be expressed as
\begin{eqnarray}
\{\J^E_{\al_i}(x),\w{\Q}^E_{-\theta+\al_i}(y)\}_{q}&=&\{
\J^E_{\al_i}(x),\w{\Q}^E_{-\theta+\al_i}(y)\}+i\ga(\al_i,-\theta+\al_i)\J^E_{\al_i}(x)\w{\Q}^E_{-\theta+\al_i}(y)
\nonumber\\
&=&2i\Bigl(N_{-\theta+\al_i,\al_i}\w{\J}^E_{-\theta+2\al_i}(x)\,\delta_{xy}+\ga(\al_i,-\theta+\al_i)\J^E_{\al_i}(x)\w{\Q}^E_{-\theta+\al_i}(y)\,\theta_{xy}\nonumber\\
&&\hspace{.5cm}+\ga(\al_i,-\theta)\J^E_{\al_i}(y)\w{\Q}^E_{-\theta+\al_i}(x)\,\theta_{yx}\Bigr).
\label{PJwQ}
\end{eqnarray}
Here we have used the identity   (\ref{id.step}).
Integrating (\ref{PJwQ}) on $x$ and $y$, the $q$-Poisson bracket between
$Q^E_{\al_i}$ and $\w{Q}^E_{-\theta+\al_i}$ is then given by
\begin{multline} \label{qwq}
\{
Q^E_{\al_i},\w{Q}^E_{-\theta+\al_i}\}_{q}
=2iD_\theta D_i^2\int_{-\infty}^\infty\!\!dx\Bigl(N_{-\theta+\al_i,\al_i}\w{\J}^E_{-\theta+2\al_i}(x) \\
+\ga(\al_i,-2\theta+\al_i)\J^E_{\al_i}(x)\int_{-\infty}^x\!\!dy\,\w{\Q}^E_{-\theta+\al_i}(y)\Bigr).
\end{multline}
Using the identity $(\al_i,-2\theta+\al_i)=-N_{-\theta+\al_i,\al_i}^2$ and
the definition \eqref{defwQ1} for $r=2$ to rewrite the right hand side of (\ref{qwq}), we obtain
\beq
\{
Q^E_{\al_i},\w{Q}^E_{-\theta+\al_i}\}_{q} = 2iN_{-\theta+\al_i,\al_i}\w{Q}^E_{-\theta+2\al_i}.
\eeq
Finally, let us evaluate the $q$-Poisson bracket between $Q^E_{\al_i}$
and $\w{Q}^E_{-\theta+ 2\al_i}$. Using the relations (\ref{PJJ}) and the
identities (\ref{id.step}) and (\ref{id.step1}), we find
\begin{eqnarray}
&&\{
\J^E_{\al_i}(x),\w{\mathfrak{Q}}^E_{-\theta+2\al_i}(y)\}_{q}\nonumber\\
&=&2i\biggl[N_{-\theta+2\al_i,\al_i}\,\w{\J}^E_{-\theta+3\al_i}(x)\,\delta_{xy}\nonumber\\
&&\hspace{1cm}+\ga(\al_i,-\theta+2\al_i)\,\J^E_{\al_i}(x)\,\w{\J}_{-\theta+2\al_i}^E(y)\,\theta_{xy}-\ga N_{-\theta+\al_i,\al_i}^2\,\J^E_{\al_i}(y)\,\w{\J}_{-\theta+2\al_i}^E(x)\,\theta_{yx}\nonumber\\
&&\hspace{1cm}-\ga^2N_{-\theta+\al_i,\al_i}\Bigl((\al_i,-\theta+\al_i)\,\J^E_{\al_i}(x)\,\J_{\al_i}^E(y)\int_{-\infty}^y\!\!dy'\,\w{\Q}^E_{-\theta+\al_i}(y')\,\theta_{xy}\nonumber\\
&&\hspace{3.5cm}+(\al_i,-\theta+\al_i)\,\J^E_{\al_i}(x)\,\J_{\al_i}^E(y)\int_{-\infty}^x\!\!dy'\,\w{\Q}^E_{-\theta+\al_i}(y')\,\theta_{yx}\nonumber\\
&&\hspace{3.5cm}+(\al_i,\al_i)\,\J^E_{\al_i}(x)\,\J_{\al_i}^E(y)\int_{-\infty}^y\!\!dy'\,\w{\Q}^E_{-\theta+\al_i}(y')\,\theta_{xy}\nonumber\\
&&\hspace{3.5cm}+(\al_i,-\theta)\,\J^E_{\al_i}(y)\,\w{\Q}^E_{-\theta+\al_i}(x)\int_{-\infty}^y\!\!dy'\,\J_{\al_i}^E(y')\,\theta_{yx}\nonumber\\
&&\hspace{3.5cm}-(\al_i,-\theta)\,\J^E_{\al_i}(y)\,\w{\Q}^E_{-\theta+\al_i}(x)\int_{-\infty}^x\!\!dy'\,\J_{\al_i}^E(y')\,\theta_{yx}\Bigr)\biggr].\label{PJQ3}
\end{eqnarray}
Integrating (\ref{PJQ3}) over $x$ and $y$,
  using the relations (\ref{rel.N-theta}) for $r=2$ and the
  definition \eqref{defwQ1} for with $r=3$, we obtain
\beq
\{
Q^E_{\al_i},\w{Q}^E_{-\theta+2\al_i}\}_{q} =2iN_{-\theta+2\al_i,\al_i}\w{Q}^E_{-\theta+3\al_i}.
\eeq
Hence we have shown that for $0\leq r \leq 2$,
\beq
\{
Q^E_{\al_i},\w{Q}^E_{-\theta+r\al_i}\}_{q} =
2iN_{-\theta+r\al_i,\al_i}\w{Q}^E_{-\theta+(r+1)\al_i}.
\label{auo}
\eeq
For the general untwisted affine Kac-Moody algebra, the $\al_i$-string through $-\theta$ has at most $\qq=2$
and $N_{-\theta+\qq\al_i,\al_i}=0$. As a consequence, the
$q$-Poisson-Serre relation (\ref{qPS1bis}) is satisfied.

\paragraph{$q$-Poisson-Serre relation \eqref{qPS2bis}.}

It follows from the $q$-Poisson bracket (\ref{qPQ1}) that the left hand side of the $q$-Poisson-Serre
relation (\ref{qPS2bis}) may be rewritten as $
2iN_{-\theta,\al_i}\{
\w{Q}^E_{-\theta+\al_i},\w{Q}^E_{-\theta}\}_{q}$. Using the Poisson brackets (\ref{PJJ}),
 the $q$-Poisson bracket for the charge densities $\w{\Q}^E_{-\theta+\al_i}(x)$
 and $\w{\Q}^E_{-\theta}(y)$ is given by
\begin{eqnarray}
\{
\w{\Q}^E_{-\theta+\al_i}(x),\w{\Q}^E_{-\theta}(y)\}_{q}&=&2i\ga\biggl[
(-\theta+\al_i,-\theta)
\,\w{\Q}^E_{-\theta+\al_i}(x)
\,\w{\J}_{-\theta}^E(y)\,\theta_{xy}\nonumber\\+
(\al_i,-\theta)&&\hspace{-.6cm}\Bigl(\w{\J}^E_{-\theta+\al_i}(x)\,\delta_{xy}-\ga N_{-\theta,\al_i}\,\J_{\al_i}^E(x)\,\w{\J}^E_{-\theta}(y)\,\theta_{xy}
\Bigr)\int_{-\infty}^x\!\!dx'\,\w{\J}^E_{-\theta}(x')\nonumber\\
+\ga N_{-\theta,\al_i} &&\hspace{-.6cm}\,\J^E_{\al_i}(x)\Bigl(
(-\theta+\al_i,-\theta)\,\w{\J}_{-\theta}^E(y)\,\theta_{xy}\int_{-\infty}^y\!\!dx'\,\w{\J}^E_{-\theta}(x')\nonumber\\
&&\hspace{2cm}+(\al_i,-\theta)\int_{-\infty}^x\!\!dx'\,\w{\J}_{-\theta}^E(x')\,\w{\J}^E_{-\theta}(y)\,\theta_{x'y}\Bigr)\biggr].
\label{qq}
\end{eqnarray}
 Integrating (\ref{qq}) over $x$ and $y$, we obtain
\begin{multline} \label{QhQ2}
\{
\w{Q}^E_{-\theta+\al_i},\w{Q}^E_{-\theta}\}_{q} = D_\theta^2D_i
(-\theta,2\al_i-\theta)
\int_{-\infty}^\infty\!\!dx\biggl[
\w{\Q}^E_{-\theta+\al_i}(x)\int_{-\infty}^x\!\!dy\,\w{\J}^E_{-\theta}(y) \\
-\ga N_{\al_i,-\theta}\,
\J^E_{\al_i}(x)
\int_{-\infty}^x\!\!dy\,\w{\J}^E_{-\theta}(y)\int_{-\infty}^y\!\!dx'\,\w{\J}^E_{-\theta}(x')\biggr].
\end{multline}
Since $(-\theta,2\al_i-\theta)=-N_{\al_i-\theta,-\theta}^2=0$,
where the last equality is because $\al_i-2\theta$ is not a root,
the $q$-Poisson bracket (\ref{QhQ2}) vanishes. This proves the
$q$-Poisson-Serre relation (\ref{qPS2bis}).

\section{Discussion}

We have shown that the conserved local charges $Q^H_{\alpha_i}$
and non-local charges $(Q^E_{\pm \alpha_i},\w{Q}^E_{\pm\theta})$ of the Yang-Baxter
$\sigma$-model satisfy the defining relations of the Poisson algebra $\mathscr U_q(L\g)$.
This result is valid when the rank of $\g$ is greater than or equal to two. We would like to conclude by
discussing a puzzle raised by this result.

\medskip

The main observation is that, despite the non-utralocal nature of the model considered, there are no ambiguities in the Poisson brackets entering the defining relations of the Poisson algebra $\mathscr U_q(L\g)$. The reason for this is that the problematic terms in derivatives of the Dirac $\delta$-distribution from the Poisson bracket of Lax matrices never showed up in the derivation. This is quite bewildering. Indeed, although the defining relations of $\mathscr U_q(L\g)$ are unambiguous, the Poisson brackets of certain conserved charges are still ill-defined! An example of this is given by the Poisson bracket between the charges $Q^E_\theta$ and $\w{Q}^E_{-\theta}$, which does not appear in the defining relations.
For concreteness, let us consider the case $\g = \mathfrak{su}(3)$.
The highest root is then $\theta=\al_1+\al_2$ ($\al_1<\al_2$). 
The charge density $\Q_\theta^E(x)$ and charge $Q_\theta^E$ are given by \cite{Delduc:2013fga}
\begin{align}
 \Q_\theta^E(x) &=\J^E_\theta(x)-\ga N_{\al_1,\al_2}\,\J^E_{\al_2}(x)
 \int_{-\infty}^x\!\!dy\,\J_{\al_1}^E(y), \label{QEthetasu3}\\
 Q_\theta^E &=D_{\al_1}D_{\al_2}\int_{-\infty}^\infty \!\!dx\,\Q^E_\theta(x).\nonumber
\end{align}
It is then clear that the Poisson bracket   $\{ \Q_\theta^E(x),
\w{\Q}^E_{-\theta}(y) \}$ contains a derivative of the Dirac $\delta$-distribution.
The value of this term follows directly from \eqref{PJJ} and reads
\beq
-8i\eta(1+\eta^2)\e^{-\ga\chi_\theta}(x)\,\e^{\ga\chi_\theta}(y)\,\partial_x\delta_{xy}. \label{ambiguity}
\eeq
As a result, the Poisson bracket $\{Q^E_\theta , \w{Q}^E_{-\theta}\}$ is not well defined. 
Note that when $\g$ is of higher rank, the expression for the charge density $\Q_\theta^E(x)$ contains further non-local terms than those in \eqref{QEthetasu3}. However, the Poisson bracket of these terms with $\w{\Q}^E_{-\theta}(y)$ does not generate any derivative of the Dirac $\delta$-distribution.
The result \eqref{ambiguity} therefore remains valid.

\medskip

Although puzzling, the situation is slightly better than in the undeformed case for the classical
analogue of the Yangian $Y(\g)$. Indeed, in order to establish the defining relations in this case one has to deal with
such ambiguities which, as already pointed out, arise even in the Poisson brackets of level $0$ charges \cite{MacKay:1992he,MacKay:2004tc,Itsios:2014vfa}.
It is interesting here to recall why, in the deformed case, not only the defining relations but in fact all the Poisson bracket relations of $\mathscr U_q(\g)$ are well defined. Indeed, all the conserved charges spanning the Poisson algebra $\mathscr U_q(\g)$ can be extracted from the monodromy matrix $T^g(\lambda)$ evaluated at the poles of the twist function. It is then immediate from \eqref{Lg} that, for these values of the spectral parameter, the Lax matrix
only depends on the field $X$. In particular, no spatial derivatives $\partial_x$ ever appear and therefore
all Poisson brackets are well defined.

\medskip

The presence of ambiguities in the full set of relations of $\mathscr U_q(L\g)$ may also be understood as follows. It was shown in
\cite{Delduc:2016ihq} that $(T^g(i\eta), T^g(-i\eta))$ satisfies the Semenov-Tian-Shansky Poisson
bracket, which corresponds precisely to the full set of Poisson bracket relations of the
Poisson algebra $\mathscr U_q(\g)$. As we have shown, enhancing the latter to an affine symmetry algebra requires working in the vicinity of the poles at $\pm i\eta$, and yet the Poisson bracket of $T^g(\lambda)$ with $T^g(\mu)$ for two arbitrary spectral parameters $\lambda$ and $\mu$ is notoriously ill-defined!

\paragraph{Acknowledgments} We thank Io Kawaguchi, Takuya Matsumoto,
Kentaroh Yoshida and Charles Young for useful discussions.  This work is partially
supported by the program PICS 6412 DIGEST of CNRS and by the French Agence
Nationale de la Recherche (ANR) under grant ANR-15-CE31-0006 DefIS.

\appendix
\section{Poisson brackets}

In this appendix we collect a number of useful Poisson brackets used in the main text.
To begin with, we shall need the Poisson brackets between the coefficients of $H^i$ and $E^{\pm\al}$
in the expressions of $X(x)$ and $\partial_xg(x) g^{-1}(x)$, \emph{i.e.} $h_i(x), e_{\pm\al}(x)$ defined in \eqref{X}
and $J_{i}^H(x), J_{\pm\al}(x)$ introduced in \eqref{Jx}. These are obtained
by comparing terms appearing on both sides of \eqref{elementarypb}. We find
\begin{eqnarray}
\{h_{i}(x),h_j(y)\}&=&0,\nonumber\\
\{h_{i}(x), {e}_{\alpha}(y)\}&=&- i\sum_{j=1}^nB_{ij}^{-1}(\al,\al_j)\,{e}_{\alpha}(x)\,\delta_{xy},\nonumber\\
\{{e}_{\al}(x),{e}_{-\beta}(y)\}&=&
\Biggl\{ \begin{array}{ll}
-4i\,\partial_x\chi_\al(x)
\,\delta_{xy}\,\delta_{\al\beta},\\
2iN_{-\beta,\al}\,e_{\al-\beta}(x)\,\delta_{xy},\hspace{4cm}\txt{if $\al\neq\beta
$},\\
\end{array}\nonumber\\
\{h_i(x),J^H_{j}(y)\}&=&B_{ij}^{-1}\,\partial_x\delta_{xy},\nonumber\\
\{h_i(x),J_{\alpha}(y)\}&=&- i\sum_{j=1}^nB_{ij}^{-1}(\al,\al_j)\,J_{\alpha}(x)\,\delta_{xy}.
\nonumber\\
\{e_{\alpha}(x),J^H_{i}(y)\}&=& i\sum_{j=1}^nB_{ij}^{-1}(\al,\al_j)\,J_{\alpha}(x)\,\delta_{xy},\nonumber\\
\{e_{\alpha}(x),J_{-\beta}(y)\}&=&
\Biggl\{ \begin{array}{ll}
\displaystyle4 i\Bigl(-\sum_{j=1}^n (\al,\al_j) J^H_{j}(x)\,\delta_{xy}-i\partial_x\delta_{xy}\Bigr)\delta_{\al\beta},
\\
2iN_{-\beta,\al}\,J_{\al-\beta}(x)\,\delta_{xy},\hspace{4cm}\txt{if $\al\neq\beta
$},\\
\end{array}\nonumber\\
\{J^H_{i}(x),J_{\beta}(y)\}&=&0,\nonumber\\
\{J_{\alpha}(x),J_{\beta}(y)\}&=&0.\nonumber
\end{eqnarray}
For simple roots $\alpha_i$ and $\alpha_j$, we then obtain
\begin{eqnarray}
\{h_{i}(x), {e}_{\pm\alpha_j}(y)\}&=&\mp i\,{e}_{\pm\alpha_j}(x)\delta_{ij}\delta_{xy},\nonumber\\
\{{e}_{\al_i}(x),{e}_{\al_j}(y)\}&=&2iN_{\al_j,\al_i}\,e_{\al_i+\al_j}(x)\,\delta_{xy},\nonumber\\
\{{e}_{\pm\alpha_i}(x),{e}_{\mp\alpha_j}(y)\}&=&\mp4 i\partial_x\chi_{\alpha_i}(x)\delta_{ij}\delta_{xy},\nonumber\\
\{e_{\pm\alpha_i}(x),J^H_{j}(y)\}&=&\pm i\,J_{\pm\alpha_j}(x)\delta_{ij}\delta_{xy},\nonumber\\
\{e_{\pm\alpha_i}(x),J_{\pm\al_j}(y)\}&=&2iN_{\al_j,\al_i}\,J_{\al_i+\al_j}(x)\,\delta_{xy},\nonumber\\
\{e_{\pm\alpha_i}(x),J_{\mp\alpha_j}(y)\}&=&4 i\Bigl(\mp\sum_{k=1}^n B_{ik} J^H_{k}(x)\delta_{xy}-i\partial_x\delta_{xy}\Bigr)\delta_{ij}.\nonumber
\end{eqnarray}
For the highest root $\theta$ and a positive root $\al\in\Phi^+$, since $\theta+\al$ is not a root, we find
\begin{eqnarray}
\{{e}_{\pm\alpha}(x),{e}_{\pm\theta}(y)\}&=&0,\nonumber\\
\{e_{\pm\alpha}(x),J_{\pm\theta}(y)\}&=&0,\nonumber\\
\{e_{\pm\theta}(x),J_{\pm\al}(y)\}&=&0.\nonumber
\end{eqnarray}
Similarly, the Poisson brackets for $\w{e}_{\pm\al}(x)$ defined in \eqref{def.we}
are computed as
\begin{eqnarray}
\{h_{i}(x), \w{e}_{\al}(y)\}&=&
-i\,\sum_{j=1}^nB_{ij}^{-1}(\al,\al_j)\,\w{e}_{
\al}(x)\,\delta_{xy},\nonumber\\
\{e_{\al}(x),\w{e}_{-\beta}(y)\}&=&
\Biggl\{ \begin{array}{ll}
\displaystyle-4i\Bigl[\partial_x\chi_\al(x)\,\delta_{xy}
-\epsilon(\al)\,2i\eta(1+\eta^2)\Bigl(\sum_{i=1}^n (\al,\al_i) J^H_{i}(x)\delta_{xy}
+i\partial_x\delta_{xy}\Bigr)\Bigr]\delta_{\al\beta},
\\
2iN_{-\beta,\al}\left[e_{\al-\beta}(x)-
\epsilon(\beta)\,2i\eta(1+\eta^2)J_{\al-\beta}(x)\right]\delta_{xy},\hspace{2.45cm}\txt{if $\al\neq\beta$,}
\\
\end{array}\nonumber
\\
\{\w{e}_{\al}(x),\w{e}_{-\beta}(y)\}&=&
\Biggl\{ \begin{array}{ll}
-4i\,\partial_x\chi_\al(x)\,\delta_{xy}\,\delta_{\al\beta},\\
2iN_{-\beta,\al}\bigl[e_{\al-\beta}(x)+(\epsilon(\al)-
\epsilon(\beta))\,2i\eta(1+\eta^2)J_{\al-\beta}(x)\bigr]\delta_{xy},\hspace{.9cm}
\txt{if $\al\neq\beta$},\\
\end{array}\nonumber
\end{eqnarray}
where $\epsilon(\al)=\txt{sign}(\al)$. We also make use of the following results
\begin{eqnarray}
\{h_{i}(x), \w{e}_{\pm\theta}(y)\}&=&\mp i\,\sum_{j=1}^nB_{ij}^{-1}(\theta,\al_j)\,\w{e}_{\pm\theta}(x)\,\delta_{xy},\nonumber
\\
\{{e}_{\pm\alpha_i}(x),\w{e}_{\pm\theta}(y)\}&=&0,\nonumber
\\
\{e_{\pm\alpha_i}(x),\w{e}_{\mp\theta}(y)\}&=&2iN_{\mp\theta,\pm\al_i}\,\w{e}_{\mp(\theta-\al_i)}(x)\,\delta_{xy},\nonumber
\\
\{e_{\theta}(x),\w{e}_{-\theta}(y)\}&=&
-4i\Bigl[\partial_x\chi_\al(x)\,\delta_{xy}
-2i\eta(1+\eta^2)\Bigl(\sum_{i=1}^n (\theta,\al_i) J^H_{i}(x)\delta_{xy}
+i\partial_x\delta_{xy}\Bigr)\Bigr],\nonumber\\
\{\w{e}_{\theta}(x),\w{e}_{-\theta}(y)\}&=&
-4 i\,\partial_x\chi_{\theta}(x)\,\delta_{xy}.
\nonumber
\end{eqnarray}
Finally, the following Poisson brackets between the densities $\J^E_\alpha(x)$ and $\w{\J}^E_\beta(x)$ hold
\begin{subequations}\label{PJJ}
\begin{eqnarray}
\{\J_{\al}^E(x),
\J^E_{\beta}(y)\}&=&2iN_{\beta,\al}\,\J_{\al+\beta}^E(x)\,\delta_{xy}\nonumber\\
&&\hspace{2cm}+i\ga(\al,\beta)\,\J_{\al}^E(x)\,\J_{\beta}^E(y)\,\epsilon_{xy},\quad \txt{if $\al+\beta\neq0$,}
\\
\{\J_{\al_i}^E(x),
\w{\J}^E_{-\theta}(y)\}&=&
2iN_{-\theta,\al_i}\,\w{\J}^E_{-\theta+\al_i}(x)\,\delta_{xy}
+i\ga(-\theta,\al_i)\,\J_{\al_i}^E(x)\,\w{\J}^E_{-\theta}(y)\,\epsilon_{xy}, \label{512a}\\
\{\J_{\al_i}^E(x),
\w{\J}^E_{-\theta+\al_i}(y)\}&=&
2iN_{-\theta+\al_i,\al_i}\,\w{\J}^E_{-\theta+2\al_i}(x)\,\delta_{xy}\nonumber\\
&&\hspace{2cm}+i\ga(-\theta+\al_i,\al_i)\,\J_{\al_i}^E(x)\,\w{\J}^E_{-\theta+\al_i}(y)\,\epsilon_{xy},
\\
\{\J_{\al_i}^E(x),
\w{\J}^E_{-\theta+2\al_i}(y)\}&=&
2iN_{-\theta+2\al_i,\al_i}\,\w{\J}^E_{-\theta+3\al_i}(x)\,\delta_{xy}\nonumber\\
&&\hspace{2cm}+i\ga(-\theta+2\al_i,\al_i)\,\J_{\al_i}^E(x)\,\w{\J}^E_{-\theta+2\al_i}(y)\,\epsilon_{xy},
\\
\{\w{\J}_{-\theta}^E(x),
\w{\J}^E_{-\theta}(y)\}&=&i\ga(-\theta,-\theta)\,\w{\J}^E_{-\theta}(x)\,\w{\J}^E_{-\theta}(y)\,\epsilon_{xy},
\\
\{\w{\J}_{-\theta}^E(x),
\w{\J}^E_{-\theta+\al_i}(y)\}&=&i\ga(-\theta,-\theta+\al_i)\,\w{\J}^E_{-\theta}(x)\,\w{\J}^E_{-\theta+\al_i}(y)\,\epsilon_{xy},\\
\{\J_{\theta}^E(x),\w{\J}^E_{-\theta}(y)\}&=&i\ga(\theta,-\theta)\,\J^E_{\theta}(x)\,\w{\J}^E_{-\theta}(y)\,\epsilon_{xy}\nonumber\\
&&\hspace{2cm}
-8i\eta(1+\eta^2)\e^{-\ga\chi_\theta}(x)\e^{\ga\chi_\theta}(y)\,\partial_x\delta_{xy}\\
&&\hspace{2cm}-4i\sum_{i=1}^n(\theta,\al_i)\left[
h_i(x)-2i\eta(1+\eta^2)\,J_{i}^H(x)\right]\delta_{xy}.\nonumber
\end{eqnarray}
\end{subequations}

\providecommand{\href}[2]{#2}\begingroup\raggedright\endgroup

\end{document}